# Automatic Event Detection in Microblogs using Incremental Machine Learning

**Tharindu Rukshan Bandaragoda**
*La Trobe Business School, La Trobe University, Victoria 3086 Australia. Email: T.Bandaragoda@latrobe.edu.au, T: +61394794700*

**Daswin De Silva**
*La Trobe Business School, La Trobe University, Victoria 3086 Australia. Email: D.DeSilva@latrobe.edu.au, T: +61394796468*

**Damminda Alahakoon**
*La Trobe Business School, La Trobe University, Victoria 3086 Australia. Email: D.Alahakoon@latrobe.edu.au, T: +6139479 3109*

The global popularity of microblogs has led to an increasing accumulation of large volumes of text data on microblogging platforms such as Twitter. These corpora are untapped resources to understand social expressions on diverse subjects. Microblog analysis aims to unlock the value of such expressions by discovering insights and events of significance hidden among swathes of text. Besides velocity; diversity of content, brevity, absence of structure and time-sensitivity are key challenges in microblog analysis. In this paper, we propose an unsupervised incremental machine learning and event detection technique to address these challenges. The proposed technique separates a microblog discussion into topics to address the key problem of diversity. It maintains a record of the evolution of each topic over time. Brevity, time-sensitivity and unstructured nature are addressed by these individual topic pathways which contribute to generate a temporal, topic-driven structure of a microblog discussion. The proposed event detection method continuously monitors these topic pathways using multiple domain-independent event indicators for events of significance. The autonomous nature of topic separation, topic pathway generation, new topic identification and event detection, appropriates the proposed technique for extensive applications in microblog analysis. We demonstrate these capabilities on tweets containing #microsoft and tweets containing #obama.[1]

## Introduction

Microblogging is becoming increasingly popular. At the time of writing there were approximately 310 million active users on Twitter and 236 million active users on Sina-Weibo. The impact of a microblog is relatively high as unlike other social networks, microblogs are visible to all users (Jansen et al., 2009). Therefore, it is important for public entities to continuously monitor this stream for relevant microblogs that indicate events of substantial impact. There have been numerous efforts in recent literature to identify such events by applying text analysis techniques; such as crises situation detection (Sakaki et al., 2010; Zhou and Chen, 2014), breaking news detection (Hu et al., 2012; Phuvipadawat & Murata, 2010), and epidemic situation detection (Culotta, 2010). In these studies, a set of keywords/hashtags (the topic) are predetermined by domain experts and the proposed technique attempts to detect bursts of these keywords/hashtags which indicate an event. Thereby, the recurrent hypothesis in these studies is that predetermined keywords/hashtags selectively identify events of interest. A clear limitation is that unforeseen events are undetected.

---

[1] This is the pre-peer reviewed version of the following article: Bandaragoda, T. R., De Silva, D., & Alahakoon, D. (2017). Automatic event detection in microblogs using incremental machine learning. Journal of the Association for Information Science and Technology, 68(10), 2394-2411., which has been published in final form at https://onlinelibrary.wiley.com/doi/abs/10.1002/asi.23896.

Hu and Liu, (2012) outlined several distinct features of text in social media text; time sensitivity, brevity, unstructured phrases. A further important feature that is often omitted is the diversity of microblogs. A microblog discussion on a multi-faceted subject would attract expressions from a broad cross-section of society. We assert that diverse discussions of this nature need to be separated into topics and each topic analysed independently for events of significance. In this research we propose to separate microblog discussions automatically into topics (topic pathways) to address this problem. The resulting topic pathways are focused on a single topic which results in more robust indication of events.

Our contributions in this paper are as follows:

- A novel unsupervised incremental learning algorithm to address the key challenges in microblog analysis. The core capabilities of this algorithms are, topic pathway separation, topic pathway evolution and incremental separation of completely new topics and subsequent topics based on already learned topics.
- A novel event detection technique that uncovers event indicators from the topic pathways generated by the above algorithm. The technique calculates the intensity of change in a pathway utilising a combination of changes in volume and positive or negative sentiment to detect potential events of interest.

We present empirical evidence of both contributions using two Twitter datasets. We demonstrate that the proposed technique can successfully separate microblog discussions into different topic pathways and also self-detect events from these pathways without requiring prior knowledge of the event.

The remainder of this paper is organised as follows: first, we report recent and relevant work on topic capture and event detection in microblog analysis; second, we present the proposed technique for topic pathway separation and event detection; third, we provide a graphic example to illustrate the proposed methods; fourth, we present the experiments and results for both datasets; and finally, the conclusion.

# Related Work

Text analytics in microblogs is more challenging than traditional text analysis due to the distinct features of microblogs. In recent research traditional text analysis techniques have been extended and new techniques developed to address those challenges. Such recent developments in the two areas relevant to this research, namely topic extraction and event detection are discussed below.

## Topic extraction

Weng et al. (2010) introduced TwitterRank, one of the earliest attempt to extract topics from a set of microblogs. It employs Latent Dirichlet Allocation or LDA (Blei et al., 2012) to identify prominent topics in a set of tweets and then determine influential authors for each topic. Hong & Davison (2010) empirically evaluated several such variants of LDA based topic models e.g., the Author-Topic model (Rosen-Zvi et al., 2004) which interoperates author information using several Twitter datasets. They argued that the effectiveness of topic models is highly influenced by the length of document and better results can be achieved by aggregating short texts of the same author into a single document.

However, Xu et al. (2011) pointed out that this strategy would only work for authors who are consistently discussing about a set of topics, and also it does not account for change in interest over time. Mehrotra et al. (2013) attempted different aggregation strategies such as aggregate tweets based on hashtags, authors and time-window (aggregate hourly tweets). They empirically showed that hashtags based aggregating is the most effective for LDA based topic extraction. However, a hashtag such as *#obama* is used to discuss a diverse set of topics, as such may result in a mix of multiple topics making it difficult to identify individual topic pathways.

Another limitation is that these topic capturing methods are designed for a static collection of microblogs and do not cater to temporal changes in a microblog discussion. LDA based topic modelling that incorporates temporality of microblog discussions was introduced in two studies (Diao and Jiang, 2013; Hong et al., 2011). However, both methods limit identification of topics to a single collection of microblogs in contrast to providing incrementally learned topic pathways over time. In addition, these methods are computationally expensive thus not scalable to handle high velocity microblog streams.

*Event detection*

An important application of microblog analytics is to capture real-world events in near real-time. Allan (2002) defines an event as a real-world occurrence with a specific time and location. However, some events cannot be constrained with a specific location as they are globally discussed. Hence, event definitions that are targeted at microblogs are often relaxed to exclude location constraints. Sakaki et al., (2010) presented one of the earliest attempts at event detection which shows that humanitarian incidents following an earthquake can be detected by tracking the frequencies of tweets that contain earthquake related phrases (e.g., earthquake, shaking) on Twitter, but a manually engineered domain specific dictionary is required to identify the relevant tweets.

Weng et al. (2011) and Xie et al. (2013) identify bursty topics as events from twitter streams by clustering bursty keywords based on their co-occurrence. Another set of approaches are proposed to cluster microblog content and monitor those clusters for temporal changes, such as number of messages, number of re-tweets and distribution of keywords in a cluster as event indicators (Abdelhaq et al., 2013; Aggarwal and Subbian, 2012; Becker et al., 2011).

The above-mentioned methods consider significant temporal changes in volume of tweets as event indicators. In contrast, Paltoglou (2015) recently introduced a sentiment based event detection technique where significant changes of sentiment can be used as event indicators with comparable results.

# The Proposed Technique

As delineated in the previous section, current methods in topic extraction are limited to a predetermined microblog discussion datasets and topics. The proposed method for topic pathway separation and event detection is motivated by the need to handle the temporal dimension of topics, generation of new topics and evolution of existing topics in microblog analytics.

The following subsections discuss topic pathway separation and the proposed event detection methods.

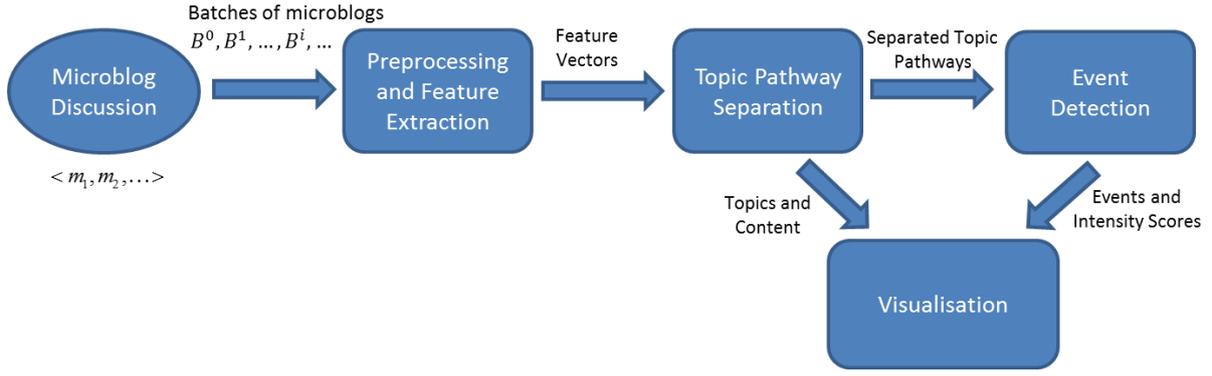

*Figure 1: The proposed technique for topic pathway separation and event detection*

As shown in Figure 1, microblog discussion $<m_1, m_2, ...>$ is first segmented into batches $B^0, B^1, ..., B^i, ...$ where each batch is a set of microblogs collected for a time period $\Delta t$.

$$B^i = \{m\}_{t=\Delta t \times i}^{t=\Delta t \times (i+1)}$$

The granularity $\Delta t$ of collecting a batch (hourly, daily, weekly etc.) can be adjusted to suit the velocity of the microblog discussion and also to match other application specific requirements. Feature vectors are extracted from each batch of microblogs and topics separated and topic pathways generated by the unsupervised learning technique. Events are detected from separated pathways using the new composite intensity value. Note that the experiments section explains pre-processing, feature extraction, and visualisation steps in more detail.

*Topic pathway separation*

A 'topic pathway' is a series of microblogs that discuss a common topic. It stretches across several batches of microblogs. A 'topic segment' is a part of the topic pathway that belongs to a particular batch. Thus, each topic pathway is a chain of topic segments.

A topic pathway $D$ can be depicted as follows:

$D = D^0 \rightarrow D^1 \rightarrow ... D^i ... \rightarrow D^N$, where a topic segment $D^i = \{m\}$ and $D^i \subset B^i$

Each topic segment is a set of microblogs that are semantically similar to each other. Hence, technically it can be considered as a clustering problem within each batch and a chain of semantically similar clusters across batches forms a topic pathway. Therefore, this problem can be reduced to building a set of semantically similar cluster chains across batches of microblogs.

The recently introduced IKASL (Incremental Knowledge Acquisition and Self Learning) algorithm (De Silva and Alahakoon, 2010) is an unsupervised incremental machine learning algorithm that can be extended to generate such topic pathways. It self-learns clusters from each data batch with generalised input from past batch based clusters for retaining past information.

*IKASL Algorithm*

In IKASL, learning occurs as a structure of layers where each layer is learned from a batch of feature vectors i.e. $n^{th}$ layer is learned from $n^{th}$ batch of feature vectors where each layer is a set of clusters representing the respective batch of feature vectors. However, clusters of layer $n$ is initialised based on the clusters formed from layer $n-1$ forming a *series of clusters*. IKASL is founded on the GSOM algorithm (Alahakoon et al., 2000), a structure adapting unsupervised clustering algorithm. The IKASL algorithm functions in three phases; initialisation, learning and generalisation. The algorithm is defined in detail below.

*Notation used in the algorithm*

$B^i$ - $i^{th}$ batch of data

$LE^i$ - $i^{th}$ learning phase

$GE^i$ - $i^{th}$ generalisation phase

$N_j$ - $j^{th}$ feature map node

$d$ - number of dimensions in the feature vector

$TE_j$ - total quantisation error of node $j$

$C^i$ - cluster representation vector learned during $GE^i$

*Initialisation phase*
- A two dimensional feature map is initialised with a network of four nodes $\{N\}$ with weight vector $w$ of size $d$.
- Initial node weights are randomly set to real values between 0 and 1.

*Learning phase*
Expands the initialised two dimensional feature map.

- A randomly selected input vector $v$ is fed to the feature map.
- The node $N_q$ closest to the input vector is identified such that: $|v - w_q| \leq |v - w|, \forall N \in \{N\}$
- The feature map adapts weights of nodes according to $v$ by modifying the weights of $\mathcal{N}_{N_q}$ (neighbourhood of $N_q$, including $N_q$)

$$w_j(k) = \begin{cases} w_j(k) + LR \times (v(k) - w_j(k)), & \text{if } N_j \in \mathcal{N}_{N_q} \\ w_j(k), & \text{otherwise} \end{cases}$$

where *LR* is the learning rate and $k = 1, 2, \ldots, d$

- $TE_{N_q} = TE_{N_q} + |v - w_q|$
- If $TE_q \geq GT$, feature map expands by linking new nodes to $N_q$

This process continues for $n$ learning iterations resulting in the feature map learning a two dimensional representation of input vectors.

*Generalisation phase*
- A *cluster representation vector* $C$ is generated by aggregating the weights of a hit node $N_h$ and the weights of its respective neighbourhood $\mathcal{N}_{N_h}$ using a fuzzy aggregation function. Hit nodes are identified by calibrating the map with input data.

*Learning from Generalisation*

After the first layer, learning phase $LE^i$ is based on the set of *cluster representation vectors* $\{C\}^{i-1}$ learned during previous generalisation phase $GE^{i-1}$. Each *cluster representation vector* becomes the weights of the starting node of a separate feature map in $LE^i$. Therefore, $|\{C\}^{i-1}|$ number of feature maps are learned in $LE^i$.

Firstly, the feature vectors are assigned to their closest *cluster representation vector* using Euclidean geometry. Those assigned feature vectors are then used to learn the respective feature maps initialised from that *cluster representation vector.*

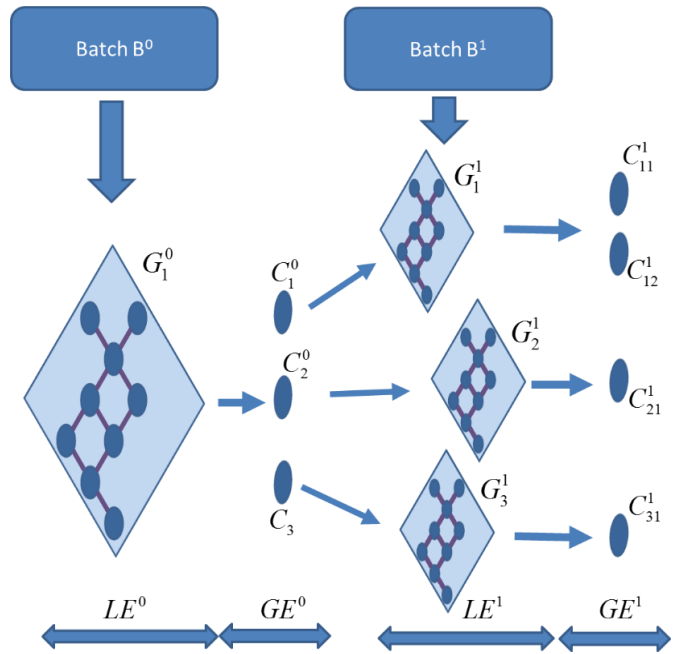

*Figure 2: Layered structure of IKASL algorithm*

Figure 2 shows the layered structure of IKASL. The feature map $G_1^0$ is randomly initialised and built using feature vectors in batch $B^0$. The cluster representations vectors $C_1^0$, $C_2^0$ and $C_3^0$ are derived from $G_1^0$. These *cluster representation vectors* are used to initialise feature maps in $LE^1$. $C_{11}^1$ and $C_{12}^1$ are derived from $G_1^1$.

*Extensions to IKASL algorithm for microblog analysis*
We extended IKASL in three ways to cater to the needs of microblog discussion as shown in Table 1.

Table 1: Requirements for microblog analysis and extensions to IKASL limitations

| Requirement | Requirement Description | Limitation | Extension |
|---|---|---|---|
| **Requirement 1** *(unstructured and dynamic vocabularies)* | A microblog discussion has a dynamic vocabulary, thus different feature spaces in different batches | Original IKASL is designed for a fixed feature space across the batches | Generate the vocabulary with the intersection of the feature spaces when comparing two vectors with different feature spaces |
| **Requirement 2a** *(improve coherence)* | Improve coherence of existing topic pathways | Original IKASL match all feature vectors to their closest cluster representation vector during learning phase | Introduced a similarity threshold when matching feature vectors to the closest cluster representation vector |
| **Requirement 2b** *(time sensitivity: identify new topic pathways)* | Identify new topic pathways to capture timely changes in a microblog discussion | | |
| **Requirement 3** *(addressing brevity)* | The aggregation function used in IKASL needs to be suitable for sparse feature spaces in microblogs | Weights of neighbouring hit nodes were aggregated by averaging to learn cluster representation vectors | Introduced *max-pooling* as the aggregation function to learn cluster representation vectors |

**Requirement 1: Handling dynamic vocabulary in microblog discussions**

Microblog users often coin new words that are not found in standard vocabulary. Mahendiran et al., (2014) pointed-out that it is difficult to define a static vocabulary as words that are used in microblogs are not known *a priori.* New words (e.g., new hashtags in Twitter) appear in relation to new events/topics. It is not viable to filter out new words as they are informative and provide better understanding about the evolution of topic pathways.

We have extended the original IKASL algorithm to handle dynamic vocabularies. As we keep the vocabulary of a single batch static, this problem is only encountered when building $LE^i$ based on *cluster representation vectors* learned from $GE^{i-1}$, as the two batches have different vocabularies used in their respective learning phases. We consider the intersection between the feature set of cluster representation vector and the feature set of the new batch, to calculate similarity between cluster representation vectors and feature vectors. For example, if the vocabulary for feature vector $v^i$ in batch $i$ be $V^i$ and vocabulary of *cluster representation vector* $C^j$ be $V^j$, cosine geometry based similarity measure is defined as:

$$Sim(C^j, v^i) = \left[ \frac{C^j \cdot v^i}{\| C^j \| \times \| v^i \|} \right]^{V^i \cap V^j} \tag{1}$$

**Requirement 2a and 2b: Improved coherence and identifying new topic pathways**

It is important to capture newly emerging topic pathways which are loosely related to existing pathways. Original IKASL algorithm links clusters to previous layers, thus all new clusters are formed having a cluster representation vector learned from previous layers as the base. When a new topic emerges, new microblogs would therefore map into an already existing topic pathway and only seen as incremental changes to an existing topic pathway.

Moreover, text datasets often face the issue of feature sparsity which is more so in microblog datasets making the nearest neighbour queries unstable (Beyer et al., 1999). Therefore, *cluster representation vectors* can become a nearest neighbour of a less similar feature vector which adversely affect the learning of feature maps and can result in less cohesive topic pathways.

In order to overcome these two issues, we introduce a similarity threshold when matching microblogs to the closest cluster representation vector. If the similarity with the closest cluster representation vector is greater than the threshold $\tau_{sim}$, then such microblogs are used to learn a new feature map $G_{new}$ that is randomly initialised as in $LE^0$ as below. Thus, for a feature vector $v$, the feature map $G$ is selected as bellow:

$$G = \begin{cases} G_j, & \text{if } C_j = \arg\max_{C \in \{C^i\}}(Sim(v,C)) \text{ and } Sim(v,C_j) > \tau_{sim} \\ G_{new}, & \text{otherwise} \end{cases} \quad (2)$$

$\tau_{sim} \in [0,1]$ is a threshold that represents the expected coherence of the identified topics. Low $\tau_{sim}$ values result in few topic pathways with less coherent content in which some might be loosely related to the particular topic. High $\tau_{sim}$ values result in more topic pathways with increased coherence.

The generation of a new feature map $G_{new}$ leads to the generation of *cluster representation vectors* that are not influenced by the previous layers and these become the basis for new topic pathways. The coherence of topic pathways is also improved by only using the feature vectors that are similar than $\tau_{sim}$ to the respective *cluster representation vector* that is used to initialise the feature map.

It should be noted that the proposed algorithm separates topic pathways autonomously. The coherence of these topic pathways can be evaluated using topic coherence measures (Röder et al., 2015). The details are further discussed in subsection *Topic Coherence* in *Experiments* section, alongside results from experiments.

**Requirement 3: Addressing the brevity issue in microblogs**

The original IKASL algorithm utilised a fuzzy aggregation function to combine learning outcomes in hit nodes and corresponding neighbourhood to generate a cluster representation vector. The feature vectors of microblogs are extremely sparse with only a handful of non-zero features in a feature vector. The nodes of feature maps learned using such features vectors have few features with significant values while other feature values are close to zero. In such situations, Boureau et al. (2010) pointed-out that max-pooling is more suitable for features that are very sparse and have a low probability of being active.

Max-pooling takes the maximum feature value for each feature from the nodes in the neighbourhood of a selected hit node. This approach does not average-out the few significant features that exist in nodes to the near zero values in other nodes thus capturing the most significant words of each node and include them in the cluster *representation vector*.

In a $d$ dimensional feature space, a *cluster representation vector* $C$ is calculated from the neighbourhood $\mathcal{N}_{N_h}$ (neighbourhood includes the hit node) of a hit node $N_h$ as follows:

$$\forall k = 1, 2, \ldots, d, \quad C(k) = \max_{N \in \mathcal{N}_{N_h}} (N(k)) \tag{3}$$

*Event detection*

The event detection technique extracts event indicators from the topic pathways. We hypothesise that if event indicators exist in topic segment $D^i$ in topic pathway $D$, then $D^i$ relates to a potential event that is related to topic of $D$.

Our proposed method linearly combines the event indicators to obtain an event score for each topic segment. If the event indicators are $I_1, I_2, \ldots, I_m (I \geq 0)$ then the event score $\mathcal{I}$ for $D^i$ is determined as:

$$\mathcal{I}(D^i) = r_1 \times I_1(D^i) + r_2 \times I_2(D^i) + \ldots + r_m \times I_m(D^i) \tag{4}$$

where $r_1, r_2, \ldots, r_m \in [0,1]$ and $\sum r = 1$

$r_i$ represents the sensitivity of an event indicator $I_i$ to the final event score. $r_i \in [0,1]$ where high or low value of $r_i$ adapts the impact of $I_i$ to the final event score accordingly. $r$ values can be empirically set based on the importance of certain indicators to suit different applications. It is also possible to learn $r$ values by training a classifier using pre-labelled records.

$\mathcal{I}$ can take values between $0$ to $+\infty$, and its value is proportionate to the significance of the event. Thus, analyst can specify an event threshold $\tau_e$ to capture significant events of interest. Also, different thresholds can be set to capture events of different significant levels.

In this paper we employed volume based and sentiment based event indicators, however any suitable event indicator can be integrated in this approach.

As pointed out in related work, volume based event indicator ($I_V$) is the most popular and in the context of a topic pathway, $I_V$ detects whether there is a significant increase in the number of microblogs in a certain topic segment indicating an increase of public interest in that topic pathways, thus a possible event. In order to capture significant changes, we set $I_V$ as the ratio between the number of messages in a topic segment and the moving average of number of messages in that topic pathway as:

$$I_V(D^i) = \frac{|D^i| \times w}{\sum_{j=i-w}^{i-1} |D^j|} \tag{5}$$

where $w$ is the window size for the moving average.

Sentiment based event indicators capture the changes of public opinion (Thelwall et al., 2011). The concept that events are associated with changes of positive and negative sentiment strength has been successfully used for event detection (Paltoglou, 2015). Sentiment analysis on microblogs is a challenging task mainly because authors of microblogs often do not follow rules of language and heavily use emoticons to express sentiment. Therefore, it is important to consider emoticons as one of the features in deriving the sentiment value.

We have investigated several state-of-the art sentiment analysis tools (Esuli et al., 2010; Socher et al., 2013; Thelwall et al., 2012), and found that SentiStrength (Thelwall et al., 2012) is well suited for our requirement; mainly because it has been specifically engineered to capture sentiment related features in microblogs. It employs a word list with sentiment strength and polarity to derive both positive and negative sentiment strengths. It considers emoticons, boosting words and negation words to strengthen or weaken the sentiment value. In addition, it considers repeated letters in a word (e.g., *sorrrry*) as an indication of adding more emphasis to the sentiment of that word.

SentiStrength provides positive and negative sentiment values for each message in a topic segment with values ranging from 1 to 4 and -1 to -4 respectively. The sentiment value of a topic segment it determined by aggregating sentiment values of microblogs belong to that topic segment and taking the average. We defined positive sentiment event indicator $I_{PS}$ and negative sentiment event indicator $I_{NS}$ in a similar manner to $I_V$ as:

$$I_{PS}(D^i) = \frac{avgPosSentiment(D^i) \times w}{\sum_{j=i-w}^{i-1} avgPosSentiment(D^j)} \quad (6)$$

$$I_{NS}(D^i) = \frac{avgNegSentiment(D^i) \times w}{\sum_{j=i-w}^{i-1} avgNegSentiment(D^j)} \quad (7)$$

where $w$ is the window size for the moving average.

With those three event indicators, event score $\mathcal{I}$ can be rewritten as follows:

$$\mathcal{I}(D^i) = r_V \times I_V(D^i) + r_{PS} \times I_{PS}(D^i) + r_{NS} \times I_{NS}(D^i) \quad (8)$$

where $r_V, r_{PS}, r_{NS} \in [0,1]$ and $\sum r = 1$.

As explained before, these $r$ values can be fine-tuned to suite the given application. For example, if the interest is on events based on negative sentiment only then $r_{NS}$ should be set to 1 while others are set to 0. Likewise, $r_V$ can be set high to capture volume based events.

## Illustrative Example

The functionality of our proposed technique for separating topic pathways and event detection is shown with an illustrative example (Figure 3).

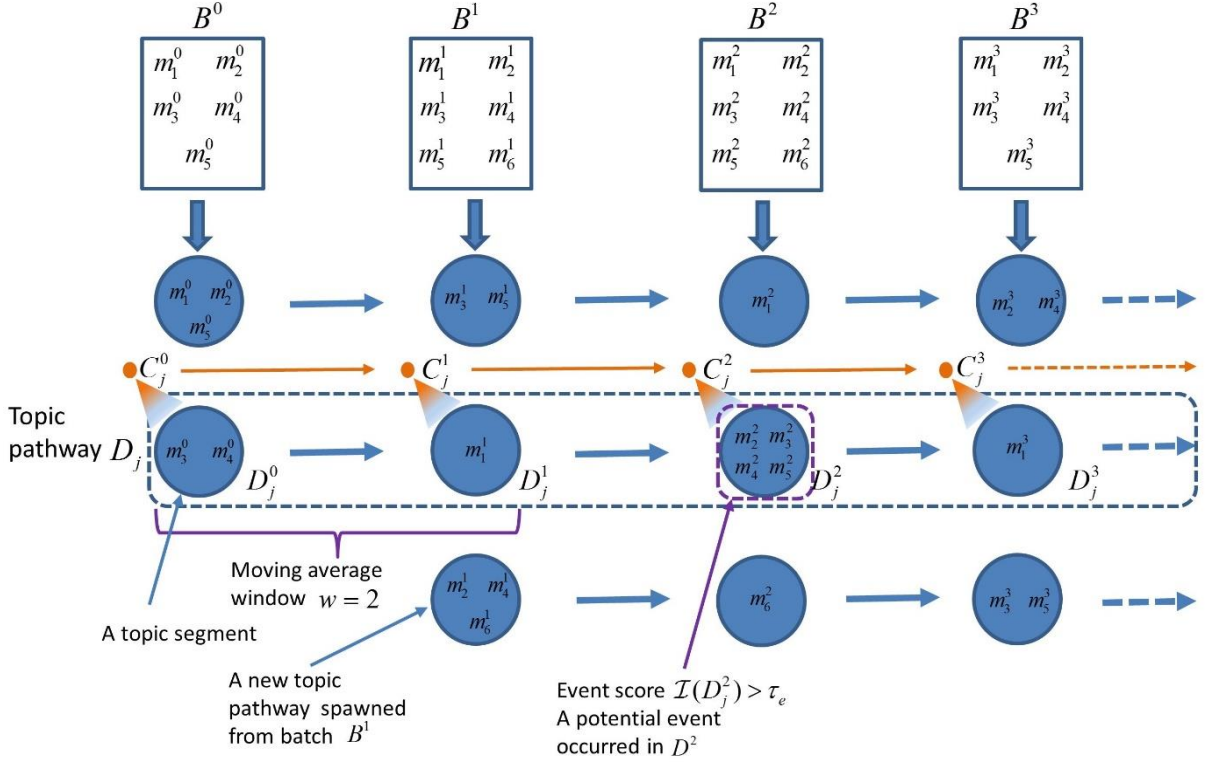

*Figure 3: An illustrative example showing topic pathway separation and event detection*

Microblogs in $B^i$ are mapped to their closest *cluster representation vector* $C \in \{C\}^i$ that are learned during the generalisation phase $GE^i$. A topic segment $D_j^i$ is formed by the set of microblogs that are mapped to the cluster representation vector $C_j^i$. As shown in Figure 3, $C_j^0$ is the closest for $m_3^0, m_4^0 \in B^0$, thus $D_j^0 = \{m_3^0, m_4^0\}$.

Since, $C_j^1$ is learned based on $C_j^0$ (the feature map that used to derive $C_j^1$ is initialised using $C_j^0$), $C_j^1$ is a timely evolution of $C_j^0$ based on $B^1$ therefore similar to each other. Similarly, $C_j^2$ is similar to $C_j^1$ and $C_j^3$ is similar to $C_j^2$. Because of this relationship among the cluster representation vectors, the topic segments formed based on these also contain semantically similar microblogs. Therefore, such topic segments are linked to form a topic pathway $D_j = D_j^0 \to D_j^1 \to D_j^2 \to D_j^3$ that contains semantically similar microblogs over time. New topic pathways can also be spawned from any batch.

Those identified topic pathways are evaluated by our proposed event detection algorithm, which collects information about event indicators and aggregate them into an event score $\mathcal{I}$. As shown, if $\mathcal{I}(D_j^2) > \tau_e$ then $D_j^2$ contains information about a potential event that is relevant to $D_j$.

The event indicators that we employed in this paper are based on detecting significant changes in volume and sentiment of a topic segment within a single pathway. For this example, the moving average window $w$ is set to 2.

# Experiments

This section demonstrates the core capabilities of the proposed algorithm. Twitter was used as the experimental microblogging platform due to its popularity and rich API. The first subsection describes the datasets and preprocessing steps. Subsequently, comprehensive accounts of the core capabilities, topic pathway separation, evolution of topic pathways, emergence of new topic pathways and automatic event detection are presented in separate subsections.

*Datasets*

**#Obama Dataset**: President Obama maintains a diverse range of public relations, both locally and internationally and he is the most followed political figure on Twitter with approximately 77.8 million followers. Thereby, analysing tweets about him is a challenging task. Separating these into topic pathways would showcase the capabilities of the proposed technique and its value over existing techniques. #Obama Dataset was collected using the keyword '#obama' for the time duration 01/12/2014 to 19/04/2015 (20 weeks).

**#Microsoft Dataset**: Microsoft Corporation has a diverse social media presence due to large portfolio of technology-centric products and services that are consumed by a variety of end-users. This diversity is a fitting testbed to further investigate the separation of a tweet stream into different topic pathways on products, services and competitors. #Microsoft Dataset was collected using the keyword '#microsoft' for the time duration 08/12/2014 to 28/06/2015 (30 weeks).

The datasets were preprocessed with duplicate removal, stop-word removal and twitter handle removal. Representative features were extracted from the pre-processed tweets employing bag-*of-words* model where terms (word or phrase) are considered as features of the tweets and feature scores are determined based on word frequency. Noun phrases from tweets were used as the features extracted using JATE toolkit (Zhang et al., 2008) omitting those less than a user-defined threshold to reduce the sparsity.

*Topic pathway separation*

This subsection demonstrates the first core capability; topic pathway separation based incremental learning outcomes.

Pathway separation occurs as the algorithm incrementally learns from the microblog stream. Some pathways remain prominent, continuing to receive significant number of tweets across each topic segment, while others lose popularity and number of tweets decreased over time.

**#Obama Dataset:** six prominent (high volume) topic pathways were identified in #Obama Dataset (labelled $TP_{Obama}^{X}$, where X is 1-6). Each topic pathway consists of a group of frequent terms that signify the topic of that pathway. These words are more frequent among the tweets of that pathway compared to the tweets of other pathways.

Table 2 presents the top 10 frequent terms of these pathways. Column 3 of this table summarises the frequent terms into a focus area.

*Table 2: Frequent terms in the prominent topic pathways of #Obama dataset*

| Topic Pathway | Frequent Terms | Key focus of the Topic Pathway |
|---|---|---|
| $TP_{Obama}^1$ | putin, russia, ukraine, economy, iran, respect, west, merkel, war, head | Relations with Russia and Ukraine |
| $TP_{Obama}^2$ | republican, party, question, tcot, congress, world, politics, cromnibu, gop, american people | Internal Politics relevant to Republican party |
| $TP_{Obama}^3$ | iran, israel, sanction, nuke, netanyahu, congress, tcot, irandeal, war, nuclear weapon, nuclear deal | Relations with Iran and Israel |
| $TP_{Obama}^4$ | nation, liberal, racism, reason, voter, democrat, consequence, tcot, gruber, pelosi | Internal Politics relevant to Democratic party |
| $TP_{Obama}^5$ | obama news, cbs news, obama video, obamacare, sore throat, washington dc, gop, castro, cnn obama, summit | News about Obama |
| $TP_{Obama}^6$ | war, afghanistan, isis, power, congress, protection siyra, iranian dissident, iraq | Terrorism, mainly ISIS |

It should be noted that some of the pathways focus on internal political issues ($TP_{Obama}^2, TP_{Obama}^4$) while others are on international political issues ($TP_{Obama}^1, TP_{Obama}^3, TP_{Obama}^5$)

In order to distinguish the salient terms that define a pathway, it is necessary to investigate their presence and prominence in other topic pathways. Table 3 tabulates this investigation. It presents the distribution of top five frequent terms of each prominent topic pathway, across the six prominent topic pathways. The rows represent the top five frequent terms of each prominent topic pathway and the columns represent the six topic pathways.

*Table 3: Distribution (%) of top five frequent terms of six prominent topic pathways*

| Topic Pathway | Frequent Terms | Distribution (%) of terms | | | | | |
|---|---|---|---|---|---|---|---|
| | | $TP^1_{Obama}$ | $TP^2_{Obama}$ | $TP^3_{Obama}$ | $TP^4_{Obama}$ | $TP^5_{Obama}$ | $TP^6_{Obama}$ |
| $TP^1_{Obama}$ | putin | 97.9 | 0.3 | 0.5 | 0.0 | 0.3 | 0.0 |
| | russia | 85.2 | 0.7 | 9.3 | 0.2 | 0.2 | 1.7 |
| | ukraine | 84.6 | 1.6 | 1.3 | 0.3 | 3.1 | 3.5 |
| | economy | 54.7 | 3.4 | 1.5 | 1.5 | 5.9 | 0.5 |
| | iran | 2.2 | 2.0 | 69.8 | 0.4 | 1.7 | 3.8 |
| $TP^2_{Obama}$ | republican | 1.0 | 81.4 | 1.5 | 2.0 | 2.8 | 0.5 |
| | party | 1.7 | 68.5 | 1.1 | 1.7 | 1.7 | 4.4 |
| | question | 3.1 | 67.3 | 4.3 | 0.0 | 0.6 | 7.4 |
| | tcot | 4.1 | 16.8 | 14.3 | 7.6 | 2.3 | 9.2 |
| | congress | 1.5 | 11.1 | 14.7 | 1.2 | 9.0 | 9.4 |
| $TP^3_{Obama}$ | iran | 2.2 | 2.0 | 69.8 | 0.4 | 1.7 | 3.8 |
| | israel | 1.8 | 3.6 | 53.9 | 1.8 | 1.7 | 2.7 |
| | sanction | 5.8 | 2.6 | 57.6 | 0.6 | 1.2 | 1.2 |
| | nuke | 2.8 | 1.2 | 70.6 | 0.0 | 0.4 | 0.4 |
| | netanyahu | 1.6 | 3.2 | 16.9 | 0.7 | 3.0 | 1.9 |
| $TP^4_{Obama}$ | nation | 0.3 | 1.6 | 0.6 | 90 | 0.9 | 0.9 |
| | liberal | 0 | 0 | 0.8 | 98.4 | 0 | 0.4 |
| | racism | 0 | 0.5 | 0 | 97.5 | 0 | 0 |
| | reason | 0.6 | 2.2 | 2.8 | 75 | 0 | 3.3 |
| | voter | 0 | 0.9 | 0 | 96.4 | 0 | 0 |
| $TP^5_{Obama}$ | obama news | 0.5 | 0.2 | 0.6 | 1 | 90.4 | 0.4 |
| | cbs news | 0 | 0 | 0.8 | 0 | 90.5 | 0.8 |
| | obama video | 2.7 | 6.7 | 0.9 | 1.3 | 49.3 | 1.8 |
| | obamacare | 2.5 | 5.7 | 1.3 | 3.2 | 69 | 1.3 |
| | washington dc | 0 | 0 | 0.9 | 11.4 | 86 | 0.9 |
| $TP^6_{Obama}$ | isis | 1.4 | 1.7 | 1.6 | 0.8 | 2 | 81.4 |
| | war | 7.6 | 2.8 | 10.7 | 2.2 | 1.6 | 56.4 |
| | iraq | 2.3 | 1.9 | 9.1 | 1.1 | 0.4 | 74.9 |
| | siyra | 7 | 19.9 | 21.6 | 0 | 0.6 | 37.4 |
| | power | 5.8 | 7.7 | 7.1 | 8.3 | 1.3 | 43.6 |

This table highlights that the top five terms of the six prominent pathways have a significant presence in the relevant pathway compared to other pathways. For, example in pathway $TP^1_{Obama}$, the top three words *putin*, *russia*, and *ukraine* have more than 80% within that pathway. The term *iran* has a low percentage in $TP^1_{Obama}$ because it has larger presence in $TP^3_{Obama}$.

Figure 4 shows the tweet volume of these topic pathways. It shows that different topic pathways peaked at different time periods. For example, in the latter part, the topic pathway $TP^3_{Obama}$ becomes popular to reflect the incidents related to the nuclear deal between USA and Iran.

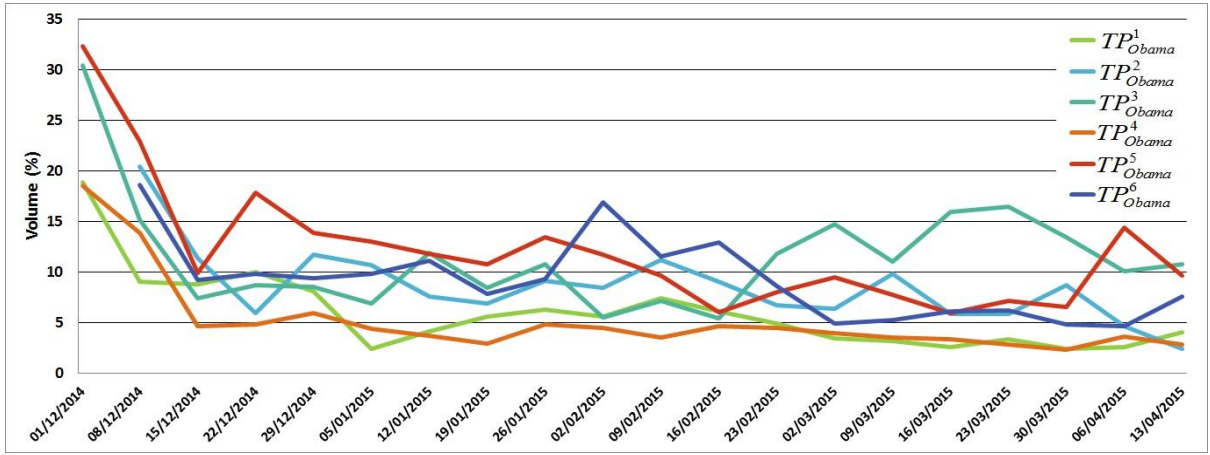

*Figure 4: Volume (%) of tweets in six prominent topic pathways of #Obama Dataset*

**#Microsoft Dataset**: Seven prominent topic pathways were identified in #Microsoft Dataset (Table 4), which were labelled using a similar approach to the above dataset.

*Table 4: Frequent terms in the prominent topic pathways of #Microsoft dataset*

| Topic Pathway | Frequent Terms | Key focus of the Topic Pathway |
|---|---|---|
| $TP^1_{Microsoft}$ | Google, android, amazon, apple, cyanogen, ibm, work, sony, app, window | Relations with Google |
| $TP^2_{Microsoft}$ | Xbox, ps4, update, windows 10, xboxone, game, Cortana, sony, bitcoin, microsoft xbox | Microsoft Xbox related products and services |
| $TP^3_{Microsoft}$ | microsoft office, ipad, android, mac, android tablet gmail, windowsphone, google, onenote, microsoft tech, office | Microsoft Office and its compatibility in different devices |
| $TP^4_{Microsoft}$ | Windows, version, windows 10, os news, future, microsoft windows, upgrade, microsoft windows10 Week, update | Microsoft Windows |
| $TP^5_{Microsoft}$ | Facebook, bing, google, microsoft bing, zdnet Apple, favour, monumental deal, search result, virtual reality | Relations with Facebook |
| $TP^6_{Microsoft}$ | windows phone, app, office, android, bitcoin, onedrive, update, bitcoin payment, tablet, ipad | Windows Phone |
| $TP^7_{Microsoft}$ | Apple, google, cloud, amazon, Samsung, microsoft store, bigdata, fight, device, patent | Relations with Apple |

Out of these seven, four topic pathways focused on products and services offered by Microsoft Corporation such as *Windows*, *Windows Phone*, *Microsoft office* and *Xbox*. Remaining three focused on relations with competitors such as with Google, Facebook and Apple. Figure 5 shows the tweet volume of these topic pathways.

The topic pathway that focuses on Windows has the highest volume in most of the weeks as Windows operating system ($TP^4_{Microsoft}$) is the most used product of Microsoft. Similarly, the topic pathway focuses on relations with Google ($TP^1_{Microsoft}$) has the highest number of tweets among the pathways that focuses on relations with competitors since both Microsoft and Google are direct competitors of several products including operations system, mobile device, search engine and virtual reality technology.

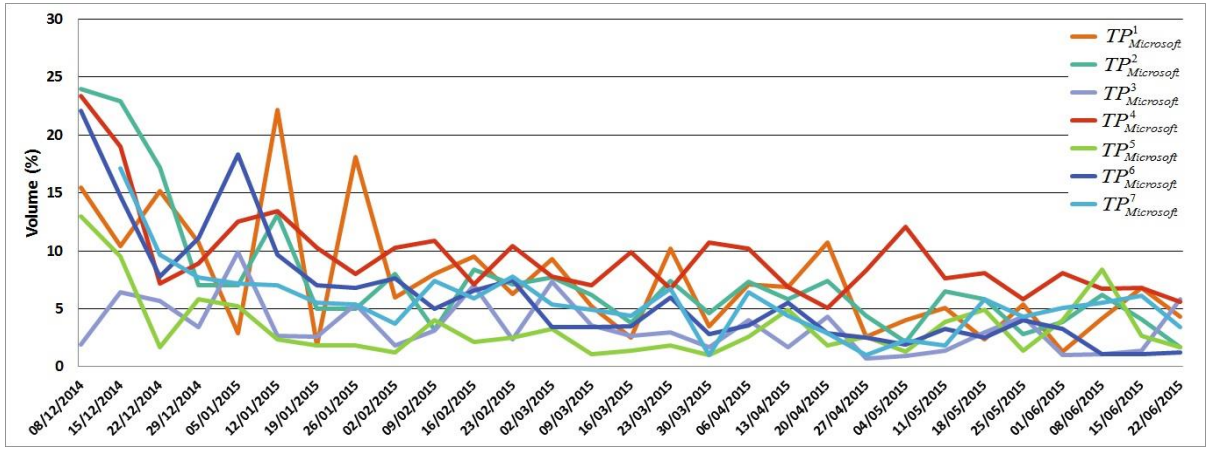

*Figure 5: Volume (%) of tweets in seven prominent topic pathways of #Microsoft Dataset*

*Topic Coherence*

It is important to evaluate the automatically identified and learned topics based on their interpretability to a human analyst. A recent research endeavour has demonstrated the use of quantitative metrics to automatically evaluate the coherence of learned topics (Röder et al., 2015). Mimno et al (2011) have demonstrated such a metric based on the principle that terms belonging to the same topic are likely to co-occur in same documents. They have empirically verified that this metric correlates with the judgement of human analysts.

This metric aggregates the term co-occurrence scores of a topic for the $n$ frequent terms of that topic. We have adapted this metric to measure and evaluate topic coherence within individual pathways. Our experiments measure the topic coherence scores of each topic for $n \in [2,100]$ and the results are presented in Figure 6 for the two datasets; (a) #Obama and (b) #Microsoft. Note that, topic coherence score calculated for the entire dataset (without pathways separation) is plotted as the baseline.

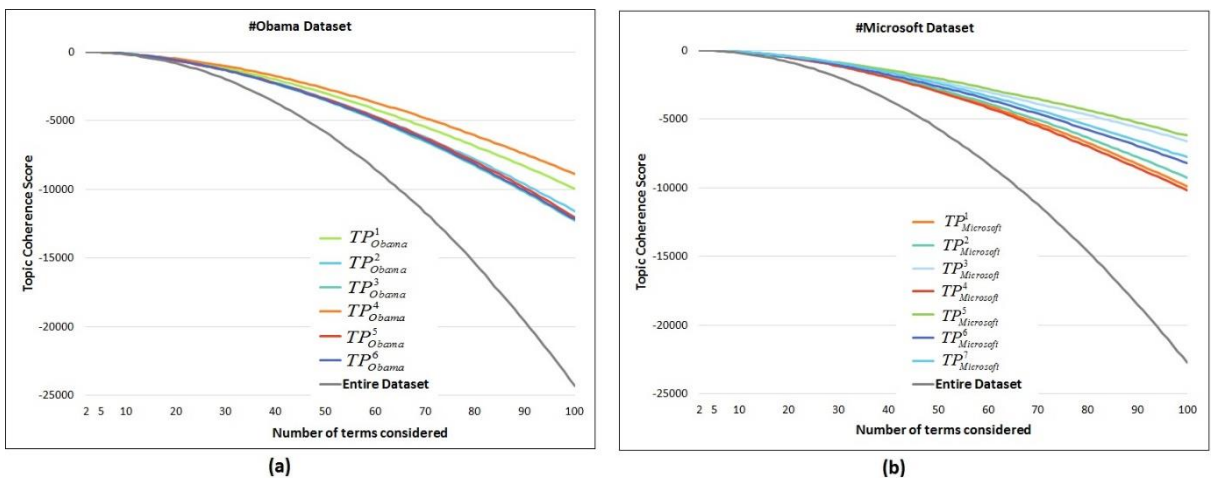

*Figure 6: Topic coherence scores for prominent topic pathways and whole dataset (for baseline) in (a) #Obama Dataset and (b) #Microsoft Dataset.*

The above results highlight that topic pathways have significant improvements in topic coherence compared to that of the entire datasets emphasising the semantic coherence of topic pathways learned by the proposed algorithm.

# Evolution of topic pathways

This subsection demonstrates the second core capability of the proposed algorithm; capturing evolution of topic pathways and temporal changes in topic segments within a topic pathway. In order to demonstrate this capability, we selected the topic pathway, $TP^1_{Microsoft}$, which captures the relationship between Microsoft and Google. It is common knowledge that both these organisations are dynamic and innovative in the technology domain, thereby microblog activity associated with both entities is equally dynamic with new subtopics (i.e., short-term topics within the particular topic pathway) being discussed. We demonstrate that the selected topic pathway $TP^1_{Microsoft}$, is capable of identifying each new subtopic as it emerges without a loss of focus on the overarching theme of Microsoft-Google relationship. Thereby, the entire topic pathway maintains focus on Microsoft-Google relationship, while each segment in the pathway focuses on different subtopics associated with this relationship as they emerge over time. The proposed algorithm's autonomous capability to capture the evolution of a topic pathway in this manner provides an analyst unique insights into short-term changes within a long-term trend. In this instance, the long-term trend is the relationship with Google and the short-term changes are the subtopics associated with both entities.

We selected nine topic segments from pathway $TP^1_{Microsoft}$ for this demonstration. Figure 7 presents word clouds for these nine topic segments. Table 5 further examines content of each of the nine topic segments. Word clouds were selected for the intuitive value in visualising fluctuations in the usage and frequency of terms. The primary observation here is that each word cloud has a unique set of frequent terms which are indicative of different subtopics.

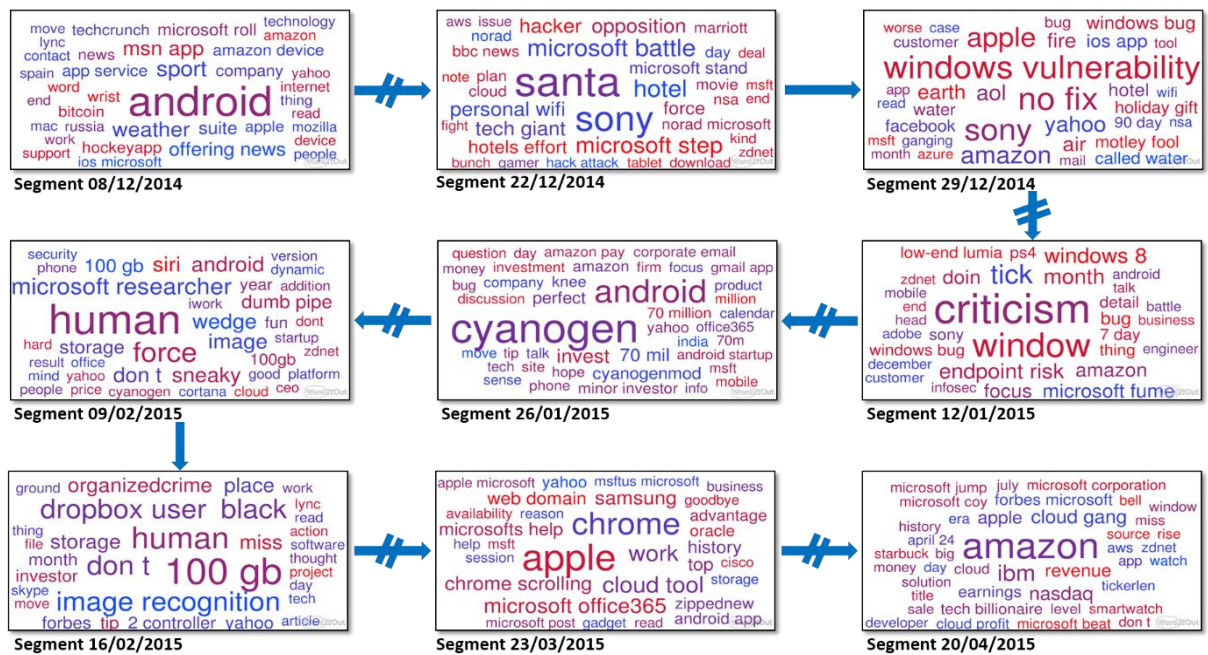

*Figure 7: Word clouds generated for nine topic segments in topic pathway, $TP^1_{Microsoft}$, which represents Microsoft-Google relationship (Most frequent terms 'Microsoft' and 'Google' were removed from word clouds for improved clarity).*

*Table 5: Further examination of topic segments illustrated as word clouds in Figure 7*

| Segment | Representative frequent terms | Subtopic(s) of the segment | Relevant News Article(s) |
|---|---|---|---|
| Segment 08/12/2014 | android, ios, msn app, sport, weather, offering news | Microsoft rolls out a set of MSN applications such as Weather, News and Sports for iOS and Android devices | MSN consumer apps arrive on Android, iOS, Amazon-CNET (11/12/2014) |
| Segment 22/12/2014 | (i) microsoft battle, personal wifi, marriott, hotel (ii) santa | (i) Google and Microsoft step in to oppose the attempts of Marriott Hotels to block personal Wi-Fi inside the hotel (ii) Google and Microsoft provides interactive features to track Santa's journey during Christmas | (i) Microsoft, Google join opposition to hotels' wi-fi blocking-ZNET(23/12/2014) (ii) Tracking Santa with help from Microsoft, Google-CNET (22/12/2014) |
| Segment 29/12/2014 | windows vulnerability, no fix, bug, windows bug | Google has openly published a Windows 8.1 vulnerability that allows users to get administrator privileges | Google discloses unpatched Windows vulnerability-ZDNET (31/12/2014) |
| Segment 12/01/2015 | criticism, windows, windows bug, microsoft fume, endpoint risk | Microsoft criticised Google for releasing information about security vulnerabilities to public | Microsoft slams Google for spilling the beans on Windows 8.1 security flaw-ZDNET(12/01/2015) |
| Segment 26/01/2015 | android, 70 mil, cyanogen, android startup, 70 million, invest | Microsoft is taking part in a $70 million investment round in a startup that is developing a competitive operating system to Android | Microsoft to Invest in CyanogenMod: What Could It Mean For Google?-Tech Times (29/01/2015) |
| Segment 09/02/2015 | human, microsoft researcher, image | Microsoft and Google are working on deep learning systems that can beat human at image recognition | Microsoft researchers say their newest deep learning system beats humans and Google-VENTUREBEAT (09/02/105) |
| Segment 16/02/2015 | (i) 100 gb, dropbox user, storage, move (ii) human, image recognition | (i) Microsoft offers 100GB cloud storage free as a promotion to get more cloud users. (ii) Microsoft and Google are working on deep learning systems that can beat human at image recognition | (i) Microsoft offers free 100GB OneDrive space to Dropbox users worldwide-ZDNET (20/02/2015) (ii) Microsoft researchers say their newest deep learning system beats humans and Google-VENTUREBEAT (09/02/2015) |
| Segment 23/03/2015 | chrome, chrome scrolling, microsofts help | Google is going to fix a scrolling issue of the Chrome browser using a Microsoft API | Google will fix Chrome's scrolling with Microsoft's help-ENGADGET(26/03/2015) |
| Segment 20/04/2015 | amazon, cloud gang, nasdaq, revenue, cloud profit | NASDAQ spiked to its highest in 15 years, propelled by the revenues from cloud services of Amazon, Microsoft and Google | Techs surge on earnings, boosts Nasdaq, S&P 500 to record close"-CNBC (24/04/2015) |

In Table 5, each row represents a topic segment. The first column is a label to identify each segment (i.e. first row in Table 5 "Segment 08/12/2014" corresponds to the first word cloud in Figure 7, also labelled "Segment 08/12/2014"). The second column denotes representative frequent terms for the corresponding segment and third column denotes the subtopic for each segment. It should be highlighted that side by side, second and third column specify the relationship between top frequent terms and the subtopic. Although the subtopics change per segment, the main theme (or focus) of the pathway remains the same. Fourth column presents a robust evaluation of the subtopic of each segment, in the form of title of corresponding news articles published in mainstream media during the same period of time.

For example, in Segment 08/12/2014, the discussion is about Microsoft providing MSN apps for android and iOS platforms. The second column presents representative frequent terms from this segment, third column presents the subtopic and fourth column validates both terms and the subtopic with the title of corresponding news article. Some representative frequent terms are common across consecutive segments which indicates that some subtopics are discussed over several weeks. For example, frequent terms in week 22/12/2014 such as 'hotel' and 'wifi' appears in the word cloud of week 29/12/2014, but with a relatively lower frequency. Similarly, 'human' and '100 gb' appears in consecutive segments 09/12/2015 and 16/02/2015.

Representative frequent terms within a segment are related to each other based on the subtopic. Representative frequent terms across segments are related to each other based on the topic pathway evolution. As noted, this topic pathway focuses on Microsoft-Google relationship, and each segment represents a different subtopic (or phase) of this relationship. Top terms across segments are able to capture this evolving topic pathway, despite differences in subtopics in each segment.

Figure 8 presents a contrasting view. It illustrates outcomes from conventional analysis of microblogs. It analyses all microblogs corresponding to the topic pathway as a single dataset. It only provides a concise view of what is discussed. Unlike Figure 7 and Table 5, it is impossible to capture evolution of a topic pathway or short-term changes in topic segments. The most frequent term (apart from 'Google') is 'android' as it is one of the main competitive product for windows operating system. Other frequent terms include 'amazon', 'ibm' and 'apple' which are competitors in the same space as Microsoft and Google. Insights gained by an analyst will be severely limited if microblog activity is explored in this manner.

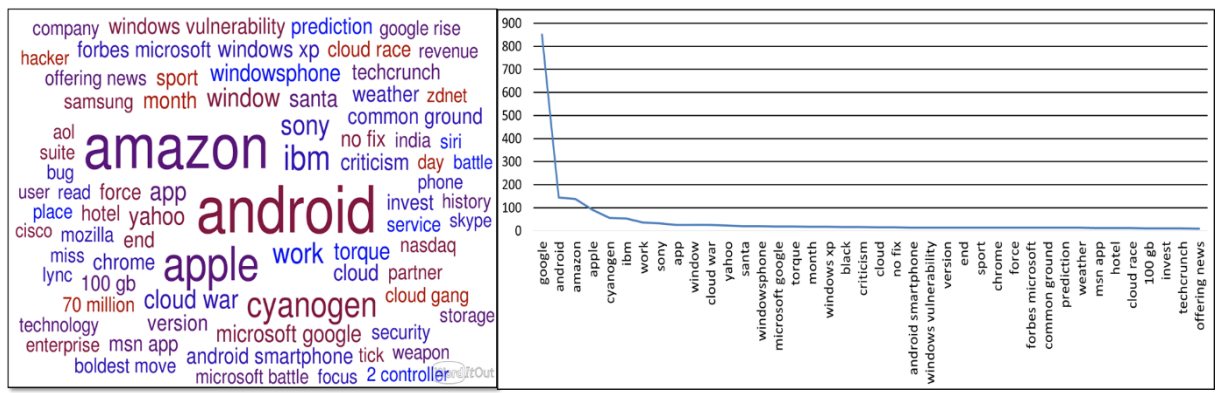

Figure 8: A word cloud and a word frequency graph generated for the topic pathway $TP^1_{Microsoft}$ (Most frequent terms 'Microsoft' and 'Google' were removed from word clouds for improved clarity).

Comparing Figure 7 with Figure 8, it is apparent that the concise view of an entire topic pathway provides limited information about the topics discussed compared to the specific views provided in topic segments. This local view helps an analyst to identify the subtopics discussed in different time periods under the same topic. Also, it is interesting to observe that some of the subtopics are public discussions about events that occurred during the time period of that topic segment. In a later subsection we demonstrate how this used for automated event detection from topic pathways.

Other topic pathways of both datasets show similar characteristics. Due to space restrictions, details of frequent terms for four prominent topic pathways (two each from two datasets) are provided in Table 6 and Table 7. Note that, the identifying terms in each pathway ('xbox' in $TP^2_{Microsoft}$; 'facebook' in $TP^5_{Microsoft}$; 'russia' and 'ukraine' in $TP^1_{Obama}$; and 'obama news' in $TP^5_{Obama}$) have been excluded to highlight subsequent frequent terms.

Table 6: Two topic pathways from #Microsoft Dataset represented using high frequent terms of the topic pathway and high frequent terms of five topic segments with highest tweet count.

| Topic pathway | Frequent terms of topic pathway | Frequent terms of five topic segments with the highest tweet count in each topic pathway | | | | |
|---|---|---|---|---|---|---|
| $TP^2_{Microsoft}$ | update, **xboxone**, game, windows 10, sony, bitcoin, plan, lumia, **ps4**, **kinect**, windows, **playstation** | *08/12/2014* **bitcoin**, windows10, november, sony, playstation | *15/12/2014* game, pandora, apple tv, app, chromecast, xboxone | *22/12/2014* **sony**, **christmas**, **ddos attack**, **playstation**, **lizardsquad** | *16/02/2015* plan, week, update, ps4, xboxone, lumia | *11/05/2015* xboxone, ps4, 10, april, playstation 4 sony |
| $TP^5_{Microsoft}$ | **bing**, **google**, **zdnet**, **apple**, favour, **monumental deal**, search result, ganging, **yahoo**, sense, future, search | *08/12/2014* **bing**, **search tool**, safe preference, monday, **graph search**, 240 million, **built-in search** | *15/12/2014* bing, google, facebooks split, zdnet key event, amber alert, motley fool | *29/12/2014* google, apple, ganging, bing, worse, microsoft service, hotmail, msn, yahoo search | *18/05/2015* google, future, lead, bing, voice communication, sweden, eurovision, winner | *08/06/2015* **monumental deal**, **virtual reality**, alibaba, sense, google, **oculus rift**, investor, gamescom |

$TP^2_{Microsoft}$ globally contains frequent words that are indicative of different products related to 'xbox' like 'xboxone', 'kinect' and 'windows 10'. Also, the word 'sony' denotes competitive manufacturer while 'ps4' and 'playstation' denotes competitive products. Notable subtopics include (i) Segment 08/12/2014: Microsoft accepts 'bitcoin' as a valid payment method to pay for Xbox games; (ii) Segment 22/12/2014: Xbox and Sony ps4 services are down due to a 'ddos attack'[2] by a hacking group called 'lizardsqsuad' on Christmas day.

$TP^5_{Microsoft}$ has frequent terms 'bing', 'znet' and 'monumental deal' that are frequent terms of different segments. In addition, there are other organisation names such as 'apple', 'google' and 'yahoo'. Notable subtopics include (i) Segment 08/12/2014: Facebook drops Microsoft Bing in favour of its own search tool; (ii) Segment 08/06/2015: Facebook and Microsoft jointly working to make the Oculus Rift virtual reality headset works with Windows 10 and Xbox.

---

[2] distributed denial-of-service attack

Table 7: Two topic pathways from #Obama Dataset represented using high frequent terms of the topic pathway and high frequent terms of five topic segments with highest tweet count.

| Topic pathway | Frequent terms of topic pathway | Frequent terms of five topic segments with the highest tweet count in each topic pathway | | | | |
|---|---|---|---|---|---|---|
| $TP^1_{Obama}$ | **putin**, economy, iran, respect, war, cuba, west, india, fight, china, weapon, **merkel**, pressure, europe **sanction**, leader | *15/12/2014* **putin**, cuba economy, **sanction**, **west**, work, crimea, problem, foxnews | *22/12/2015* economy, dow, gas, foxnews, putin, cuba, head, golf, hawaii | *02/02/2015* putin, wwiii, merkel, kiev, arms, fight, Moscow, weapon, china, Kerry, respect | *09/02/2015* **putin**, weapon, war, merkel, poroshenko, fighting, hollande, **minksummit** | *16/02/2015* putin, merkel, hollande, respect, allies, weapon, sanction, shot, war, west |
| $TP^5_{Obama}$ | **cbs news**, obama video, obamacare sore throat, gop, cuba, castro, acid reflux, summit, congress, prayer breakfast, iran, hospital, seattle, **cnn**, **abc news** | *22/12/2014* troop, 6 year, hawaii, apron, hacking row, cbs news, obamacare, afghanistan | *05/01/2015* **community college**, france, seattle, **malia obama**, **union tour**, congress, **housing move** | *12/01/2015* obamacare, **cameron**, **methane emission**, iran, **middle class**, obama video | *02/02/2015* slavery, prayer breakfast, isis, crusade, obamas, comparison christianity | *06/04/2015* castro, cuba, summit, stage, saturday, panama, abc news, historic meeting |

$TP^2_{Obama}$ contains opinions about Obama handling Russia and Ukraine issues. Frequent terms of this pathway contains the names such as 'putin' (President of Russia), and 'merkel' (Chancellor of Germany). Also, the words 'sanction' is frequent showing that people talk about sanctions on Russia. Subtopics include (i) 15/12/2014: Obama's support for new U.S. sanctions to Russia; (ii) 09/02/2015: ahead of Minsk summit, Obama threatened the Russian President with serious consequences for involvement in the Ukrainian conflict.

$TP^5_{Obama}$ contains news tweets about Obama often tagged as '#obama #news'. Tweets posted by news agencies have tags used by them such as 'cbs news', 'cnn' and 'abc news'. Topic segments in this pathway is often a mixture of subtopics about different news articles. Notable subtopics: (i) 05/01/2015: Obama proposes free two years at community colleges, Malia Obama supports anti-cop rap group, Obama to tout Arizona housing initiative; (ii) 12/01/2015: Obama to call for tax increase on rich to help middle class, proposes reduction in methane emissions from oil and gas, and Obama and Cameron talk counter terrorism and economy.

## Emergence of new topic pathways

This subsection demonstrates the third core capability of the proposed algorithm; the detection of emerging topic pathways in the twitter stream. These new topic pathways are loosely related to previously identified pathways. Such new topics can be internal (e.g., new product or service launch) or external (e.g., interaction with a new organisation) to the business entity. Subsequently, some of them fade away as people lose interest, while some topic pathways prevail due to continuing interest.

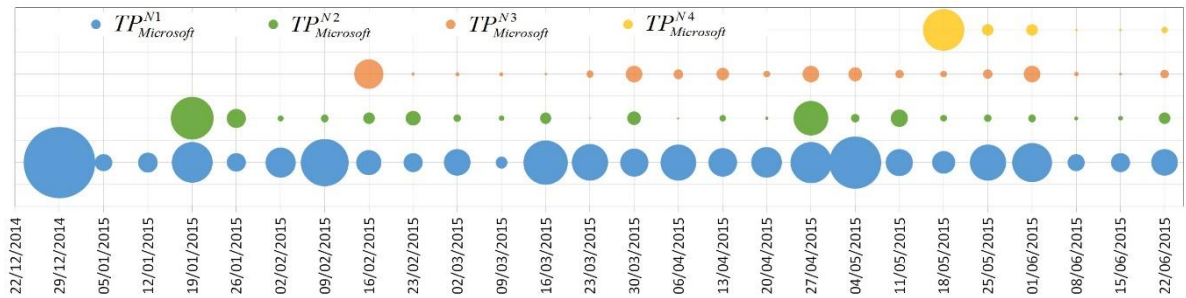

*Figure 9: Four new topic pathways emerged in #Microsoft Dataset. Each topic pathway is represented using a chain of bubbles where each bubble is a topic segment. Diameter of a bubble is set to number of tweets in the topic segment.*

Figure 9 shows selected four new topic pathways that emerged in different weeks in the #Microsoft Dataset.

$TP_{Microsoft}^{N1}$ : (frequent terms- *windows 10, internet explorer, version, pirate, upgrade*) initiation coincides with the week where Microsoft announces that there will be a new web browser in Windows 10 operating system. Since then it has significant amount of tweets that mainly discuss about new features in Windows 10. It is interesting to note that this pathway was separated as a new topic despite having a topic pathway focused on Windows) in general. We hypothesis that it is due to differentiating terms (e.g., windows 10, release, future, upgrade). Unlike other new topic pathways, this pathway retains its popularity throughout the time span of the dataset.

$TP_{Microsoft}^{N2}$ : (frequent terms- *microsoft hololen, world, hologram, future, google glass*) initiated when Microsoft introduced its virtual reality glasses 'Hololens' during 'Microsoft Consumer Preview' on 21/01/2015. After the initial peak it declined to few tweets a week except for the week 27/04/2015 in which the capabilities of Hololens was demonstrated at Microsoft Build Developer Conference.

$TP_{Microsoft}^{N3}$ : (frequent terms- *linux, service, windows, python, bigdata*) is initiated when Microsoft announces the incorporation of Linux operating system in its big data services provided in Microsoft Azure cloud platform. This move was widely discussed as Linux is a direct competitor to Windows.

$TP_{Microsoft}^{N4}$ : (frequent terms- *salesforce, talk, msft, price, 55b*) is focused on Microsoft's attempted acquisition of the company 'Salesforce'. It created a hype on twitter but declined in popularity when it failed.

Figure 10 presents the topic coherence metric for the above mentioned topic pathways plotted alongside the seven original topic pathways of the #Microsoft dataset. It shows that new topic pathways (except $TP_{Microsoft}^{N1}$) have increased topic coherence scores than the other prominent pathways. This is because they are focused on topics which emerged as a result of particular events and thus contain more closely related terms that are uniquely used to discuss them. $TP_{Microsoft}^{N1}$ has frequent terms shared with the topic pathway focused on *Windows* in general ($TP_{Microsoft}^{4}$) which results in relatively low coherence scores.

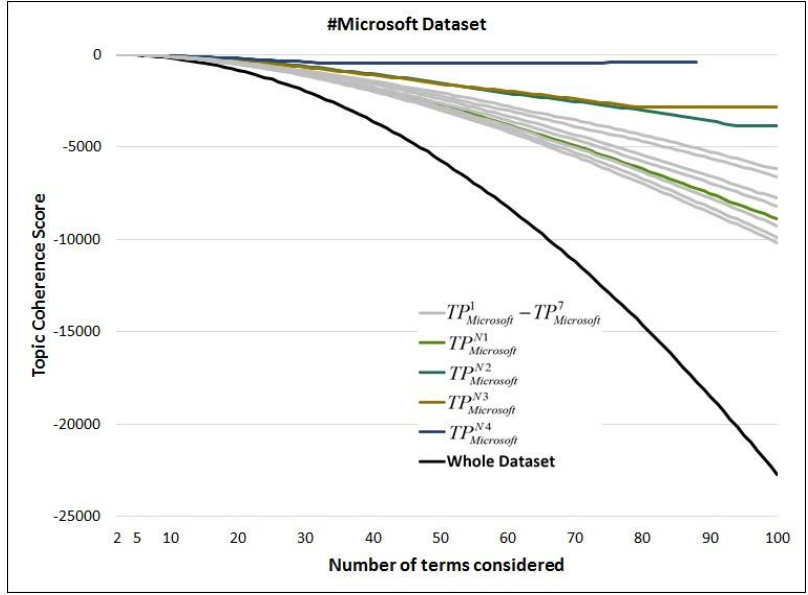

*Figure 10: Topic coherence of newly emerged topic pathways*

Detection of these new topic pathways highlights the capability of our proposed algorithm to identify previously unseen topics at any stage of the Microblog stream. This capability is due to the second extension of the extended IKASL algorithm, which facilitates learning a new randomly initialised feature map from microblogs that are less similar to the existing topic pathways when there is a burst of microblogs on a new topic.

## Automatic event detection

The aim of this section is to demonstrate the fourth capability; automatic detection of events from topic pathways using event indicators.

As previously shown in Equation 8, event score is determined by combining three event indicators for each topic segment in topic pathways.

The $r$ parameters in Equation 8 decides the impact of each event indicator to the final event score. However, we have observed that the frequency and intensity changes of volume is higher compared to the changes of positive and negative sentiment. In order to compensate for this behaviour $r_V$ is set to a lower value than others as follows $r_V = 0.1$, $r_{PS} = 0.45$ and $r_{NS} = 0.45$. We also employed $w = 2$ for the calculations of $I_V$, $I_{PS}$ and $I_{NS}$. $\tau_e$ is set to 1.0 for this experiment and we filtered-out events that belong to topic segments containing less than 1% of the tweets in that batch.

Tables 8 and 9 present the top 10 events identified based on event score $\mathcal{I}$ in the two data sets. The frequent terms that are related to an event is derived by capturing the frequent terms that appear in the topic segment containing the event, but does not appear in the frequent terms of the $w$ previous topic segments of the same pathway. We analysed online news articles from major news agencies to verify the significance of events detected by the algorithm.

*Table 8: Top 10 events detected based on event score $\mathcal{I}$ for #Microsoft Dataset*

| Week | $\mathcal{I}$ | Event Indicator Scores | | | Topic pathway | Frequent terms related to the event | Verification evidence based on news articles |
| --- | --- | --- | --- | --- | --- | --- | --- |
| | | $0.1 \times I_V$ | $0.45 \times I_{PS}$ | $0.45 \times I_{NS}$ | | | |
| 27/04/2015 | 1.62 | **0.69** | 0.53 | 0.40 | $TP_{Microsoft}^{N2}$ | microsoft hololen, google glass, build2015, augmented reality, future | Microsoft demonstrated the capabilities of Hololens in Microsoft Build Developer Conference |
| 22/12/2014 | 1.37 | 0.10 | 0.52 | **0.75** | $TP_{Microsoft}^{2}$ | psn, ddos attack, christmas, lizardsquad, credit, lizard squad | Xbox Live online services were down on Christmas day after a DDoS attack by a group called 'LizardSquad'. |
| 16/03/2015 | 1.31 | **0.52** | 0.36 | 0.43 | $TP_{Microsoft}^{N1}$ | windows 10, pirated version, summer, pirate, password | Microsoft announced that it will give free upgrades to Microsoft 10, even for pirate copies |
| 30/03/2015 | 1.26 | **0.48** | 0.42 | 0.36 | - | Iphone, rogue window, android phone, office lens, document, powerful scanner | Microsoft launches Office Lens scanner app to iPhone and Android for the first time |
| 12/01/2015 | 1.22 | 0.18 | **0.54** | 0.50 | $TP_{Microsoft}^{1}$ | criticism, window, tick, endpoint risk, windows 8 | Google released malicious code which can be used to exploit Microsoft Windows |
| 11/05/2015 | 1.16 | 0.20 | 0.40 | **0.56** | - | microsoft hyperlapse, video, desktop, hyperlapse, shaky video, launch | Microsoft launches Hyperlapse, a video smoothing software for Android and Windows phones |
| 30/03/2015 | 1.15 | 0.15 | **0.52** | 0.48 | $TP_{Microsoft}^{4}$ | open source, windows sensor, open-source, microsoft exec, sale | Microsoft executive says that it is possible that Windows become open-source in future |
| 09/02/2015 | 1.14 | 0.24 | 0.40 | **0.50** | $TP_{Microsoft}^{7}$ | google app, note, iwork, windows user, microsoft office, 700 billion, watch, google threaten, work | Google just threatened to expose Apple and Microsoft security flaws. Apple opens up iWork to Windows users for free |
| 09/03/2015 | 1.13 | 0.30 | **0.45** | 0.38 | - | microsofts cortana, report, standalone app, ios device, digital assistant, siri | Microsoft announced its intelligent personal assistant Cortana will be released for iPhone and Android |
| 09/02/2015 | 1.13 | 0.20 | **0.48** | 0.45 | $TP_{Microsoft}^{N1}$ | public preview, lumia, first preview, microsoft windows10forphone, lumia 635 | Microsoft delivers first public preview of Windows 10 for phones |

*Table 9: Top 10 events detected based on Event score $\mathcal{I}$ for #Obama Dataset*

| Week | $\mathcal{I}$ | Event Indicator Scores | | | Topic pathway | Frequent terms related to the event | Verification evidence based on news articles |
|---|---|---|---|---|---|---|---|
| | | 0.1 $\times I_V$ | 0.45 $\times I_{PS}$ | 0.45 $\times I_{NS}$ | | | |
| 19/01 /2015 | 1.22 | 0.33 | **0.52** | 0.37 | - | tuesday, netanyahu, india visit, proof, nice, world, obamainindia | Obama visits India. |
| 05/01 /2015 | 1.22 | 0.10 | 0.42 | **0.70** | $TP^3_{Obama}$ | gitmo, france, freedom, paris, press, terror, war | Obama released 9 suspects held in GTMO prison. Obama condemns Charlie Hebdo attack saying terrorists fear freedom |
| 26/01 /2015 | 1.16 | 0.08 | 0.43 | **0.65** | $TP^6_{Obama}$ | white house, taliban, muslim province, whitehouse, japanese journalist, islamicstate, beheading | ISIS threatens to behead #Obama and transform America into a Muslim province. ISIS killed second Japanese Hostage |
| 30/03/ 2015 | 1.16 | 0.23 | **0.49** | 0.44 | $TP^3_{Obama}$ | irandeal, lying, obama kerry, detail, framework, trust, irantalk, good deal, nuke agreement | |
| 02/02 /2015 | 1.15 | 0.12 | 0.41 | **0.62** | $TP^6_{Obama}$ | death, january, crusade, unemployment rate, campaign, christianity, muslim leader, teaparty, immigration | Obama, at National Prayer Breakfast, compares ISIS to violence of Christian Crusades. USA unemployment rate rises to 5.7% in January |
| 19/01/ 2015 | 1.13 | 0.07 | **0.57** | 0.49 | - | battle, marijuana legalization, tntweeters obama, crisis, obama cybercare | Obama on Thursday said he expects more states to experiment with marijuana legalization |
| 16/02 /2015 | 1.13 | 0.14 | **0.60** | 0.39 | - | fabric, founding, woven, love, obama islam, giuliani, obamas love, gov walker, mayor giuliani | Obama created controversy by saying "Islam Has Been Woven Into the Fabric of Our Country Since Its Founding". New York Mayor Rudy Giuliani said Obama doesn't love America |
| 22/12 /2014 | 1.13 | 0.15 | **0.50** | 0.48 | $TP^5_{Obama}$ | 6 year, apron, hacking row, novice, obama motorcade, afghanistan, trend, malaysian leader, afghan milestone | Obama hails end of combat operations in Afghanistan. |
| 23/03/ 2015 | 1.12 | 0.10 | 0.49 | **0.53** | $TP^5_{Obama}$ | obamacare, price, double funding, antibiotic resistance, national plan, senate, superbug | Obama announced a 5-year action plan to prevent lives lost from antibiotic resistance. |
| 19/01/ 2015 | 1.12 | 0.19 | **0.57** | 0.36 | $TP^1_{Obama}$ | respect, russia, ukraine, saudi arabia, pressure, saudi king, putin, india trip, russian aggression | Obama cuts short presidential trip to India, and travelled to Saudi Arabia to pay respects to King Abdullah. Obama pledges more pressure on Russia as Ukraine clashes broaden. |

The above tabulated results demonstrate and verify the authenticity of events detected by the proposed algorithm. It is interesting to note that these events were captured exclusively based on social expressions on Twitter. It is also interesting to observe that different events were triggered by high values from different indicators. Volume based indicators are high in the first event of Table 8, which is about an event where Microsoft demonstrated the capabilities of its virtual reality product Hololens. As shown in Figure 9 a new topic pathway is spawned when Hololens was first introduced by Microsoft on 21/01/2015.

Another example where $I_v$ is high is the third event in Table 8, which is due to Microsoft's statement that Windows 10 would be a free upgrade for all users (including those with pirated). This statement attracts attention on social media as it was a surprise move by Microsoft (quite often aggressive about software piracy). This peak of volume can also be seen in Figure 9 for $TP^{N1}_{Microsoft}$.

There are several events in both tables that are significantly high in negative sentiment event indicator score. Event 2 in Table 8 is the highest triggered by the event where Xbox Live online services were offline due to a hacker attack on Christmas day. This event has caused issues for online gaming community who took it into Twitter criticising Microsoft for lack of security and sluggish response.

Event 2 in Table 9 is highly negative, which was due to firstly, people condemning the release of nine suspects held in Guantanamo Bay and secondly the Charlie Hebdo Shooting in the same week. Social commentary combined both incidents stating that releasing of terror suspects can lead to more attacks. This incident is identified from topic pathway focused on Iran/Israel relations, which is loosely related to the event since most of the released suspects are from Middle East.

Events with high positive sentiment are comparatively rare. This is largely because social media is more often used to express negative sentiment more than positive. It also supports the finding that popular events are often associated with negative sentiment (Thelwall et al., 2011).

This subsection aptly demonstrates the final capability of the proposed algorithm, automated event detection from microblog streams. The incorporation of several event indicators allow us to capture events across diverse domains without having to develop tailored event capture mechanisms for individual applications.

## Discussion and Conclusion

In this paper, we presented a novel technique to address the key challenges of diversity, brevity, absence of structure and time-sensitivity in microblog analysis and the automatic detection of events of significance.

The ability of IKASL algorithm to self-learn from the data and generate separate topic pathways addresses the problem of diversity. The capacity to incrementally learn with new data while maintaining an account of the past addresses time sensitivity. The issues due to brevity and unstructured nature in microblogs is addressed by introducing a dynamic vocabulary for new incoming terms and an extended thresholding technique.

The topic pathways contain microblog content focused on a single topic, which reduces noise due to diversity of content and improves detection of events. We have also shown that combining multiple event indicators yields a robust event detector that could be deployed without the need of customising to specific domains or different data volumes.

As demonstrated, topic pathway separations are self-learned without domain specific knowledge and also completely data-driven without any prior knowledge about the number of topics, compared to traditional topic modelling techniques. In contrast to having a static set of topics, topic pathways could be considered as temporal account of topics which is a chain of topic segments. We demonstrated that these topic segments provide information about the subtopics which appear across different time periods. Our method also identifies new topics as the microblog discussion progresses over time and spawns separate pathways to represent such topics.

The proposed event detection method captures significant changes in topic segments compared to the previous segments in the same topic pathway. Such changes were monitored using an intensity value made up of three event indicators volume, positive and negative sentiment. The results show that different event indicators have contributed to the high intensity values in different situations and as such justify the use of a composite intensity value made up of multiple indicators.

The proposed approach delivered promising results on the two datasets tested. As future work we recommend further testing on datasets from diverse application domains as well as from different types of social media platforms. Further work can look into the sensitivity of volume and sentiment based event detectors to the final event score; and how it changes in diverse application domains.

## Acknowledgments

The authors would like to thank the anonymous reviewers and edition for their valuable comments and suggestions to improve the quality of this paper. This work was supported by an Australian Government Research Training Program Scholarship. Authors would also like to acknowledge the financial and in-kind support from the Data to Decisions Cooperative Research Centre (D2D CRC) as part of their analytics and decision support program.

## References

Abdelhaq, H., Sengstock, C., & Gertz, M. (2013). EvenTweet: Online Localized Event Detection from Twitter. *Proceedings of the VLDB Endowment*, *6*(12), 1326–1329.

Aggarwal, C., & Subbian, K. (2012). Event Detection in Social Streams. In *Proceedings of the 2012 SIAM International Conference on Data Mining* (Vol. 12, pp. 624–635).

Alahakoon, D., Halgamuge, S. K., & Srinivasan, B. (2000). Dynamic Self-Organizing Maps with Controlled Growth for Knoledge Discovery. *IEEE Transactions on Neural Networks*, *11*(3), 601--614.

Becker, H., Naaman, M., & Gravano, L. (2011). Beyond Trending Topics: Real-World Event Identification on Twitter. In *Proceedings of the Fifth International Conference on Weblogs and Social Media (ICWSM)* (pp. 1–17).

Beyer, K., Goldstein, J., Ramakrishnan, R., Shaft, U., Beeri, C., & Buneman, P. (1999). When Is "Nearest Neighbor" Meaningful? *Database Theory—ICDT'99*, *1540*, 217–235.

Blei, D. M., Ng, A. Y., & Jordan, M. I. (2012). Latent Dirichlet Allocation. *Journal of Machine Learning Research*, *3*(4–5), 993–1022.

Boureau, Y.-L., Ponce, J., & LeCun, Y. (2010). A Theoretical Analysis of Feature Pooling in Visual Recognition. In *Proceedings of the 27th international conference on machine learning (ICML-10)* (pp. 111–118).

Culotta, A. (2010). Towards detecting influenza epidemics by analyzing Twitter messages. In *Proceedings of the First Workshop on Social Media Analytics - SOMA '10* (pp. 115–122).

De Silva, D., & Alahakoon, D. (2010). Incremental knowledge acquisition and self learning from text. In *Proceedings of the International Joint Conference on Neural Networks (IJCNN)* (pp. 1--8).

Diao, Q., & Jiang, J. (2013). A Unified Model for Topics , Events and Users on Twitter. In *Proceedings of the 2013 Conference on Empirical Methods in Natural Language Processing* (pp. 1869–1879). Association for Computational Linguistics.

Esuli, A., Sebastiani, F., & Moruzzi, V. G. (2010). SentiWordNet 3.0: An Enhanced Lexical Resource for Sentiment Analysis and Opinion Mining. In *In Proceedings of the Seventh Conference on International Language Resources and Evaluation (LREC'10)* (Vol. 10, pp. 2200-2204).

Hong, L., & Davison, B. (2010). Empirical study of topic modeling in twitter. *Proceedings of the First Workshop on Social Media Analytics*, 80–88.

Hong, L., Dom, B., Gurumurthy, S., & Tsioutsiouliklis, K. (2011). A time-dependent topic model for multiple text streams. In *Proceedings of the 17th ACM SIGKDD international conference on Knowledge discovery and data mining - KDD '11* (pp. 832–840).

Hu, M., Liu, S., Wei, F., Wu, Y., Stasko, J., & Ma, K.-L. (2012). Breaking news on twitter. In *Proceedings*

*of the SIGCHI Conference on Human Factors in Computing Systems* (pp. 2751–2754).

Hu, X., & Liu, H. (2012). Text analytics in social media. In C. C. Aggarwal & C. Zhai (Eds.), *Mining Text Data* (pp. 385–414).

James Allan. (2012). *Topic Detection and Tracking - Event-based Information Organization*. Springer Science & Business Media.

Jansen, B. J., Zhang, M., Sobel, K., & Chowdury, A. (2009). Twitter power: Tweets as electronic word of mouth. *Journal of the American Society for Information Science and Technology*, *60*(11), 2169–2188.

Mahendiran, A., Wang, W., Arredondo, J., Lira, S., Huang, B., Getoor, L., … Ramakrishnan, N. (2014). Discovering Evolving Political Vocabulary in Social Media. In *Behavior, Economic and Social Computing (BESC), 2014 International Conference on* (pp. 1–7).

Mehrotra, R., Sanner, S., Buntine, W., & Xie, L. (2013). Improving LDA topic models for microblogs via tweet pooling and automatic labeling. In *Proceedings of the 36th international ACM SIGIR conference on Research and development in information retrieval- SIGIR '13* (pp. 889-892).

Mimno, D., Wallach, H. M., Talley, E., Leenders, M., & McCallum, A. (2011). Optimizing semantic coherence in topic models. *Proceedings of the Conference on Empirical Methods in Natural Language Processing*, 262–272.

Paltoglou, G. (2015). Sentiment-Based Event Detection in Twitter. *Journal of the Association for Information Science and Technology*, 1576-1587.

Phuvipadawat, S., & Murata, T. (2010). Breaking news detection and tracking in Twitter. *Proceedings - 2010 IEEE/WIC/ACM International Conference on Web Intelligence and Intelligent Agent Technology - Workshops.*

Röder, M., Both, A., & Hinneburg, A. (2015). Exploring the Space of Topic Coherence Measures. In *Proceedings of the Eighth ACM International Conference on Web Search and Data Mining - WSDM '15* (pp. 399–408).

Rosen-Zvi, M., Griffiths, T., Steyvers, M., & Smyth, P. (2004). The author-topic model for authors and documents. In *Proceedings of the 20th conference on Uncertainty in artificial intelligence* (pp. 487–494).

Sakaki, T., Okazaki, M., & Matsuo, Y. (2010). Earthquake Shakes Twitter Users: Real-time Event Detection by Social Sensors. In *Proceedings of the 19th international conference on World wide web (WWW '10)* (pp. 851–860).

Socher, R., Perelygin, A., & Wu, J. (2013). Recursive Deep Models for Semantic Compositionality Over a Sentiment Treebank. *Proceedings of the Conference on Empirical Methods in Natural Language Processing (EMNLP)*, *1631*, 1631–1642.

Thelwall, M., Buckley, K., & Paltoglou, G. (2011). Sentiment in Twitter Events. *Journal of the American Society for Information Science and Technology*, *62*(2), 406–418.

Thelwall, M., Buckley, K., & Paltoglou, G. (2012). Sentiment Strength Detection for the Social Web. *Journal of the American Society for Information Science and Technology*, *63*(1), 163–173.

Weng, J., Lim, E.-P., Jiang, J., & He, Q. (2010). TwitterRank: Finding Topic-sensitive Influential Twitterers. In *Proceedings of the third ACM international conference on Web search and data mining* (p. 261).

Weng, J., Yao, Y., Leonardi, E., Lee, F., & Lee, B. (2011). Event Detection in Twitter. *ICWSM*, *11*(98),

401–408.

Xie, W., Zhu, F., Jiang, J., Lim, E.-P., & Wang, K. (2013). TopicSketch: Real-Time Bursty Topic Detection from Twitter. In *2013 IEEE 13th International Conference on Data Mining* (pp. 837–846).

Xu, Z., Ru, L., Xiang, L., & Yang, Q. (2011). Discovering User Interest on Twitter with a Modified Author-Topic Model. In *2011 IEEE/WIC/ACM International Conferences on Web Intelligence and Intelligent Agent Technology* (Vol. 1, pp. 422–429).

Zhang, Z., Iria, J., Brewster, C., & Ciravegna, F. (2008). A comparative evaluation of term recognition algorithms. In *Proceedings of the Sixth International Conference on Language Resources and Evaluation (LREC08)*.

Zhou, X., & Chen, L. (2014). Event detection over twitter social media streams. *VLDB Journal*, *23*(3), 381–400.